\documentstyle[preprint,aps,epsf,floats]{revtex}
\begin{document}

\begin{titlepage}

\preprint{KUNS 1452}

\title{AMD + HF model and its application to Be isotopes}

\author{Akinobu Dot$\acute{\rm e}$ \and and  Hisashi Horiuchi}
\address{Department of Physics, Kyoto University, Kyoto 606-01, Japan}

\author{Yoshiko Kanada-En'yo}
\address{Institute of Particle and Nuclear Studies,
High Energy Accelerator Research Organization, Midori-cho 3-2-1, 
Tanashi, Tokyo188, Japan}

\maketitle

\begin{abstract}
In order to study light unstable nuclei systematically, we propose a 
new method ''AMD + Hartree-Fock''. This method introduces the concept
of the single particle orbits into the usual AMD. Applying AMD + HF to 
Be isotopes, it is found that 
the calculated lowest intrinsic states with plus and minus parities 
have rather good correspondence with the explanation by the 
two-center shell model. In addition, by active
use of the single particle orbits extracted
from AMD wave function, we construct the first excited 0$^+$ state of 
$^{10}$Be. The
obtained state appears in the vicinity of the lowest 1$^-$ state. This 
result is consistent with the experimental data.

\end{abstract}

\end{titlepage}

\section{Introduction}

\vspace{0.5cm}

For the study of the structure of light unstable nuclei
which are and have been extensively studied experimentally by the use
of radioactive beams [1-3],
various
theories [4-20]
 have been used [21,22]
. Among those theories, 
the antisymmetrized molecular dynamics (AMD)
approach has already proved to be useful and successful\CITE{AMD}. 
One of the large merits of the AMD approach is that it does not rely 
on any model 
assumptions such as axial symmetry of the deformation, existence of the 
clustering and so on.

Until now this AMD approach has not explicitly utilized the concept of 
the single particle motion in the mean field. The experimental
data, however, show that very often we can get better understanding of
the structure of unstable nuclei in terms of the single particle
motion in the mean field. In order to see this point in more detail, we
here discuss briefly some features of Be isotopes as an example. 
In Be isotopes there is the famous problem that the ground state 
$^{11}$Be has an anomalous parity. According to the usual shell model, 
its spin-parity 
should be $\frac{1}{2}^-$. But in experiments it is $\frac{1}{2}^+$. 
An answer to this problem is
that due to the deformation the $\Omega^\pi$=$\frac{1}{2}^+$ level in sd shell
comes down below the upper $\Omega^\pi$=$\frac{1}{2}^-$ and the last 
neutron occupies 
the lowered $\Omega^\pi$=$\frac{1}{2}^+$. 
$^{11}$Be is also famous as a  halo nucleus. Since the lowered orbit 
contains s-orbit component, we can understand that the halo property is 
due to the long tail of the s-orbit. 
For the convenience of later discussion we call this lowered
 $\Omega^\pi$=$\frac{1}{2}^+$ level '' halo level''.

\begin{figure}[h]
\epsfxsize=0.6 \textwidth
\centerline{\epsffile{ 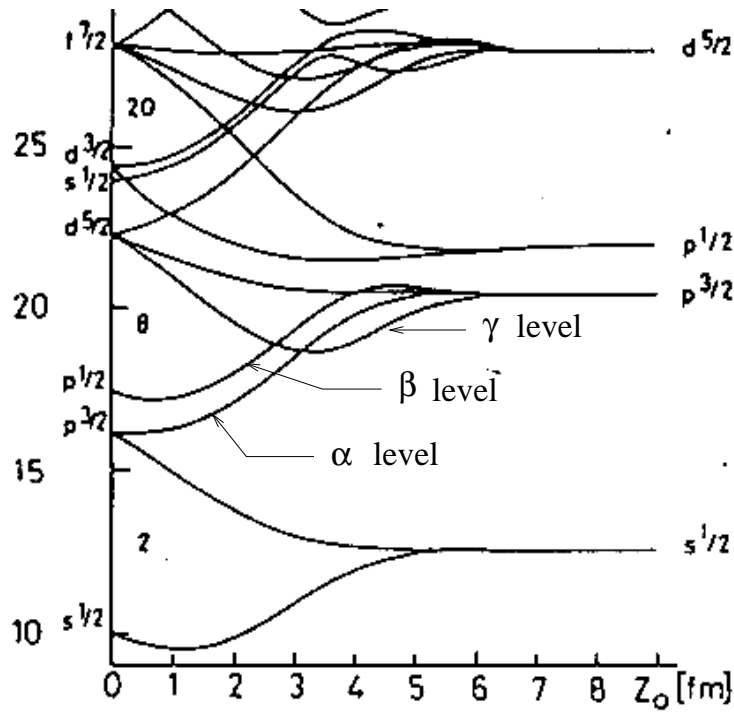}}
\caption{ 
The complete neutron level scheme of the two-center shell
model as a function of the eccentricity parameter $Z_0$. On the
ordinate at the left are given the spherical shell model states of the 
original nucleus together with the energy scale, while on the
right-hand side the quantum numbers of the spherical shell model
states in the fragment nuclei are seen. This figure is quoted from [24].}
\end{figure}

In order to discuss other Be isotopes than $^{11}$Be, 
we use the single particle diagram by 
the two-center shell model[18-20], which is shown in FIG.1\CITE{tcsm}. 
According to the AMD calculation for Be isotopes, we can support the
idea that Be isotopes have the core part with an approximate dumbbell 
structure of two $\alpha$ clusters, which is the reason why we use the
two-center shell model. 
$^9$Be is known to have three rotational bands which have the band head states
with J$^\pi$=$\frac{3}{2}^-$, $\frac{1}{2}^-$ and $\frac{1}{2}^+$.  
We can explain these three bands in the two-center shell model  
as follows. We put the four neutrons in order from the lowest 
level at about Z$_0$=2.5 fm in this model, and when the last neutron is put 
in the levels of $\Omega^\pi$=$\frac{3}{2}^-$($\alpha$ level), 
$\frac{1}{2}^-$($\beta$ level) and $\frac{1}{2}^+$($\gamma$ level)
, J$^\pi$ becomes $\frac{3}{2}^-$, 
$\frac{1}{2}^-$ and $\frac{1}{2}^+$, respectively. 
Here, Z$_0$ is the eccentricity parameter, namely the distance between 
two centers in FIG.1. 
$^{10}$Be is known to have two rotational bands which have the band head states
with J$^\pi$=0$^+$ and 1$^-$. In the
same way as above, the four neutrons are put in order from the lowest
level at about Z$_0$=3 fm and when the configurations of the last two
neutrons 
are $[\Omega^\pi=\frac{3}{2}^-]^2$ and $[\Omega^\pi=\frac{3}{2}^-]^1 
[\Omega^\pi=\frac{1}{2}^+]^1$, J$^\pi$ becomes 0$^+$ and 1$^-$, 
respectively. 
Here we note that the $\Omega^\pi$=
$\frac{1}{2}^+$ level is the halo level which appeared in $^{11}$Be. 

These arguments show that by using the single particle orbits and 
by constructing the excited state by particle-hole excitation 
we can study the structure of Be isotopes systematically. 
Therefore we recognize 
the importance of the single particle motion in the mean field. 

In this paper we propose the ''AMD + Hartree-Fock'' method (AMD+HF)
which introduces the concept of the single particle orbits 
into the usual AMD. 
We extract the component of the mean field from the AMD wave
function. Once we get the single particle wave 
functions in the mean field, we
can study the structure of the excited states by constructing various
particle-hole configurations. When we need elaborate description of
the single particle motion like in the case of neutron halo
phenomena, we need to improve the AMD single particle wave function in
such a way that we adopt the superposition of several Gaussians
instead of the single Gaussian wave packet in the usual AMD.

This paper is composed of six sections. In the second section we
explain the formalism of AMD+HF method. In the third section we show
the single particle levels of the 
ground states of Be isotopes which are calculated with AMD+HF. 
In the forth section we show the single particle levels obtained with
AMD+HF as a function of the deformation parameter. 
In the fifth section, based on the motivation of the 
AMD+HF, which means active use of the single particle levels, we
construct the first plus-parity excited state of $^{10}$Be by putting
two neutrons into the halo 
level and study it. 
And the last section is for the discussion and the summary.

\section{AMD+HF method}

\subsection{Variation under the constraint of deformation}

\subsubsection{Hamiltonian and wave function}

In this paper, we used the Hamiltonian and wave function as described below.

The Hamiltonian has the form:
\begin{eqnarray*}
\hat{H} & = & \hat{T}+\hat{V}_c+\hat{V}_{LS}-\hat{T}_G . \\
\end{eqnarray*}
Here $\hat{T}$, $\hat{V}_c$, $\hat{V}_{LS}$ and $\hat{T}_G$ stand for
the kinetic energy, the central force, the LS force and the
center-of-mass kinetic energy, respectively. We neglected the Coulomb
force. we used the Volkov No.1 force\CITE{volk} as the central force and
the G3RS force\CITE{G3RS} for the LS force. These forces have the
following forms, 
\begin{eqnarray*}
\hat{V}_c & = & \frac{1}{2}\sum_{i,j}
(w+b\hat{P}_\sigma+h\hat{P}_\tau-m\hat{P}_\sigma\hat{P}_\tau)
[V_Re^{-({\hat{\bf r}}_{ij}/r_R)^2}+V_Ae^{-({\hat{\bf
r}}_{ij}/r_A)^2}] ,\\
\hat{V}_{LS} & = & \frac{1}{2}\sum_{i,j} V_{LS}[e^{-({\hat{\bf
r}}_{ij}/r_{LSR})^2}-e^{-({\hat{\bf
r}}_{ij}/r_{LSA})^2}]P(^3O)\hat{\bf L}_{ij}(\hat{\bf s}_i+\hat{\bf
s}_j) ,
\end{eqnarray*}
where m=0.56, w=1$-$m, b=h=0, V$_R$=144.86 MeV, V$_A$=$-$83.34 MeV, r$_R$=0.82
fm, r$_A$=1.6 fm, V$_{LS}$=900 MeV, r$_{LSR}$=0.20 fm, r$_{LSA}$=0.36 
fm. P($^3$O) is the projection operator onto 
the triplet odd state.

The wave function of total system is expressed by a Slater determinant:
\begin{eqnarray*}
|\Phi> & = & \det[|\varphi_i(j)>] ,\\
<{\bf r}|\varphi_i> & = & (\frac{2\nu}{\pi})^{3/4}
\sum_\alpha C_i^\alpha \exp[-\nu({\bf r}-\frac{{\bf
Z}_i^\alpha}{\sqrt{\nu}})^2 \; |\beta_i> ,
\end{eqnarray*}
where $|\beta_i>$ is the spin-isospin function. 
In AMD+HF, single particle wave function is in general a linear
combination of several Gaussians in order to describe single particle state
more adequately than usual AMD. However in this paper we represented
the single particle state by a single Gaussian wave packet for
simplicity. $|\Phi>$ is projected to parity eigenstates $|\Phi^\pm>$
and the energy variation is made after parity projection. 
\begin{eqnarray*}
|\Phi^\pm> & = & \frac{1}{\sqrt{2}}[|\Phi>\pm {\cal P}|\Phi>] .
\end{eqnarray*}

\subsubsection{Cooling equation with constraint}
 
As mentioned above, the wave function is parameterized by 
$\{C^\alpha_i,{\bf Z}^\alpha_i\}$. These parameters are determined 
by solving the following cooling equation, 
\begin{eqnarray*}
\dot{X}_i & = & (\lambda+i\mu)\frac {1}{i \hbar} 
\frac{\partial {\cal H}}{\partial X_i^*}  \hspace{0.5cm} {\rm and} 
\hspace{0.3cm} C.C. , \\
{\cal H} & = & \frac{<\Phi^\pm|\hat{H}|\Phi^\pm>}{<\Phi^\pm|\Phi^\pm>} 
.
\end{eqnarray*}
Here $\{X_i\}$ means $\{C^\alpha_i,{\bf Z}^\alpha_i\}$, $\lambda$ is an 
arbitrary real number and $\mu$ is a 
negative arbitrary number. 
We can easily prove that the energy of the total system decreases
with time, 
\begin{eqnarray*}
\frac{d}{dt}{\cal H} & < & 0 .
\end{eqnarray*}

We often need to obtain the minimum-energy state under some
 condition. For example, the condition is that the 
center-of-mass is fixed to the coordinate origin, or that the
deformation parameter is fixed to some value, etc. 
In such case, the condition is combined to the cooling equation
by the analogy of the Lagrange-multiplier method. When the condition 
is represented as
${\cal W}$(X$_i^\alpha$,X$_i^{\alpha*}$)=0, the previous cooling 
equation is changed as below:
\begin{eqnarray}
\dot{X}_i & = & (\lambda+i\mu)\frac {1}{i \hbar} [\frac{\partial {\cal H}}
{\partial X_\imath^*}+\eta\frac{\partial {\cal W}}
{\partial X_\imath^*}] \hspace{0.5cm} {\rm and} 
\hspace{0.3cm} C.C. ,
\end{eqnarray}
where $\eta$ is a Lagrange-multiplier function, which is determined by 
$\frac{d}{dt} {\cal W}=0$:
\begin{eqnarray*}
0 & = & \frac{d}{dt}{\cal W} \\
  & = & \sum_i[\frac{\partial{\cal W}}{\partial X_\imath} \dot{X}_i+
\frac{\partial{\cal W}}{\partial X_\imath^*} \dot{X}_i^*] \\
  & = & \sum_i[\frac{\partial{\cal W}}{\partial X_\imath} \{
(\lambda+i\mu)\frac {1}{i \hbar} (\frac{\partial {\cal H}}
{\partial X_\imath^*}+\eta\frac{\partial {\cal W}}{\partial
X_\imath^*})\} \\
  &  & \mbox{}+
\frac{\partial{\cal W}}{\partial X_\imath^*}\{(\lambda-i\mu)
\frac {-1}{i \hbar} (\frac{\partial {\cal H}}
{\partial X_\imath}+\eta\frac{\partial {\cal W}}
{\partial X_\imath})\}] .
\end{eqnarray*}
From this equation we get
\begin{eqnarray*}
\eta & = & -\frac{\cal F}{\cal G} , \\
{\cal F} & = & \sum_i [(1-i\frac{\lambda}{\mu})\frac{\partial {\cal
W}}{\partial X_\imath} \frac{\partial {\cal H}}{\partial X_\imath^*}+
(1+i\frac{\lambda}{\mu})\frac{\partial {\cal
W}}{\partial X_\imath^*} \frac{\partial {\cal H}}{\partial X_\imath}]
, \\
{\cal G} & = & 2\sum_i \frac{\partial {\cal W}}{\partial X_\imath}
\frac{\partial {\cal W}}{\partial X_\imath^*} .
\end{eqnarray*}
One may think that this method is applied only to the case where the number of 
conditions is one. But
we can use it also in the case where several conditions exist. 
When several conditions are represented as 
\begin{eqnarray*}
{\cal W}_1=0, {\cal W}_2=0, \cdots, {\cal W}_n=0 ,
\end{eqnarray*}
these conditions can be represented by a single equation,
\begin{eqnarray*}
{\cal W}=c_1|{\cal W}_1|^2+ \cdots +c_n|{\cal W}_n|^2 = 0 .
\end{eqnarray*}
Here $c_1 \cdots c_n$ are positive coefficients which adjust the difference 
of scale between ${\cal W}_1, {\cal W}_2, \cdots, {\cal W}_n$. 
By this way, we can always use Eq.(1) without introducing several 
Lagrange-multipliers. 

\subsubsection{Deformation parameter}

The deformation parameter ($\beta$,$\gamma$) can be obtained from the
following relations, 
\begin{eqnarray*}
A_x \equiv \frac{<x^2>^{\frac{1}{2}}}{[<x^2><y^2><z^2>]^\frac{1}{6}}
& = & \exp[\sqrt{\frac{5}{4 \pi}} \beta \cos(\gamma + \frac{2 \pi}{3})]
 ,\\
A_y \equiv \frac{<y^2>^{\frac{1}{2}}}{[<x^2><y^2><z^2>]^\frac{1}{6}}
& = & \exp[\sqrt{\frac{5}{4 \pi}} \beta \cos(\gamma - \frac{2 \pi}{3})]
 ,\\
A_z \equiv \frac{<z^2>^{\frac{1}{2}}}{[<x^2><y^2><z^2>]^\frac{1}{6}}
& = & \exp[\sqrt{\frac{5}{4 \pi}} \beta \cos \gamma ] ,
\end{eqnarray*}
where 
\begin{eqnarray*}
<x^2> & = & \frac{<\Phi| \frac{1}{A} \sum_i x_i^2 |\Phi>}{<\Phi|\Phi>} 
, \hspace{0.5cm} {\rm etc .}
\end{eqnarray*}
Here x, y, z directions are the directions of the principal axes of
inertia which are the coordinate axes of the body-fixed frame. 
We can easily confirm that this definition of deformation parameter is 
the same as that of Bohr model to the first order. Here expectation value of 
every operator is calculated with the intrinsic wave function 
before parity projection. In addition $<x^2>$, $<y^2>$ and $<z^2>$ is 
redefined so that $<z^2> \geq <y^2> \geq <x^2>$.
We must note the fact that in this definition z axis is not always
the axial symmetric axis of the system. In the case of prolate
 deformation, 
z axis determined by this definition is certainly the axial symmetric 
axis, 
while in the case of oblate deformation it is not the axial symmetric 
axis. 

When we execute the cooling calculation with deformation constraint, 
it consumes time that in every step of the cooling we calculate the 
inertial axes to determine the deformation parameters. 
Therefore we use rotational invariant scalar quantities for
calculating the deformation parameters so as to avoid the
time-consuming procedure. 
We define the tensor quantity {\em Q} in any coordinate system as below:
\begin{eqnarray*}
Q_{ij} & = & <r_i r_j> ,
\end{eqnarray*}
where i, j are x, y, z.
Then calculating the trace of the tensors {\em Q} and {\em QQ}, 
we get the rotational invariant quantities Tr{\em Q} and Tr{\em QQ}. 
Since calculating these quantities in the space-fixed frame is identical to 
that in body-fixed frame, we obtain the following relations:
\begin{eqnarray*}
TrQ_{space-fixed} & = & TrQ_{body-fixed} \\
 & = & \sum_i <r^2_i>_{body-fixed} = [<x^2><y^2><z^2>]^\frac{1}{3}
\sum_i A_i^2 \\
 & =  & [<x^2><y^2><z^2>]^\frac{1}{3} \cdot (3+\frac{15}{4 \pi} \beta^2
+\cdots) ,\\
TrQQ_{space-fixed} & = & TrQQ_{body-fixed} \\
 & = & \sum_i <r^2_i>^2_{body-fixed} = [<x^2><y^2><z^2>]^\frac{2}{3}
\sum_i A_i^4 \\
 & =  & [<x^2><y^2><z^2>]^\frac{2}{3} \cdot (3+\frac{15}{\pi} \beta^2
+\cdots) .
\end{eqnarray*}
Using these quantities, we define {\em D} as below:
\begin{eqnarray*}
D & \equiv & \frac{TrQQ}{(TrQ)^2} \hspace{1cm}{\rm in}
 \hspace{0.2cm}{\rm space-fixed \; frame} .
\end{eqnarray*}
Then we can calculate the deformation parameter $\beta$ as
\begin{eqnarray*}
\beta & \simeq & \sqrt{\frac{2\pi}{5}(3D-1)} .
\end{eqnarray*}
Thus we can obtain $\beta$ without using the inertial axes. 
This approximation is enough correct to the second order of $\beta$, 
which in ordinary case is less than unity. 

When we impose the constraint that $\beta$ is fixed to some $\beta_0$, 
we use {\em D} and {\em D}$_0$ corresponding to $\beta$ and $\beta_0$ in the
cooling calculation. 
The constraint condition function ${\cal W}_D$ is given as 
\begin{eqnarray*}
{\cal W}_D & = & (D-D_0)^2 .
\end{eqnarray*}

\subsection{Single particle orbits described by AMD}

\subsubsection{Method of calculating single particle orbits}

As mentioned in the introduction, we need to calculate the single particle 
orbits. We calculate them by the method shown below. This method extracts 
the single particle orbits from AMD wave function by mimicking the
Hartree-Fock theory. 

In the Hartree-Fock method, 
the matrix elements h$_{ij}$ of the single particle Hamiltonian 
in orthonormal base $\{|\phi_i>\}$ are given as
\begin{eqnarray*}
h_{ij} & = & <\phi_i|\hat{t}|\phi_j> + \sum_k <\phi_i \phi_k|\hat{v}
(|\phi_j \phi_k>-|\phi_k \phi_j>) .
\end{eqnarray*}
By diagonalizing this h$_{ij}$, we obtain the set of the single 
particle wave functions and the single particle energies. 
Now we construct Hartree-Fock-like single particle orbits and levels 
from total wave function of AMD. 
The single particle wave functions $\{\varphi_i\}$ of AMD are not
 orthonormal.  
So first we construct the orthonormal base by making
linear-combination of $\{\varphi_i\}$. We calculate the set of eigenvalues
and eigenvectors of the overlap matrix {\em B},
\begin{eqnarray*}
\sum_j B_{ij} c^\alpha_j & = & \mu^\alpha c^\alpha_i , \\
B_{ij} & \equiv & <\varphi_i|\varphi_j> .
\end{eqnarray*}
Eigenvectors $\{c^\alpha_i\}$ are normalized. We define new base 
$\{|f^\alpha>\}$ as follows:
\begin{eqnarray*}
|f_\alpha> & = & \frac{1}{\sqrt{\mu^\alpha}} \sum_i c^\alpha_i
|\varphi_i> .
\end{eqnarray*}
We can confirm easily that this set forms an orthonormal base.

Using this base $|f^\alpha>$ we construct the matrix $\{h_{\alpha
\beta}\}$ corresponding to the single particle Hamiltonian of
Hartree-Fock as follows:
\begin{eqnarray*}
h_{\alpha \beta} & = & <f_\alpha|\hat{t}|f_\beta> + 
\sum_\gamma <f_\alpha f_\gamma|\hat{v}
(|f_\beta f_\gamma>-|f_\gamma f_\beta>) .
\end{eqnarray*}
Diagonalizing this matrix $\{h_{\alpha\beta}\}$, we calculate the eigenvalues 
$\{\epsilon^p\}$, and eigenvectors $\{g^p_\alpha\}$,
\begin{eqnarray*}
\sum_\beta h_{\alpha \beta} g^p_\beta & = & \epsilon^p g^p_\alpha .
\end{eqnarray*}
The eigenvectors $\{g^p_\alpha\}$ are normalized.
The single particle wave function $|p>$ belonging to the single particle
energy $\epsilon^p$ is given as 
\begin{eqnarray*}
|p> & = & \sum_\alpha g_\alpha^p |f^\alpha> \\
    & = & \sum_\alpha g_\alpha^p [ \frac{1}{\sqrt{\mu^\alpha}} 
          \sum_i c^\alpha_i |\varphi_i>] \\
    & = & \sum_i [\sum_\alpha g_\alpha^p
          \frac{1}{\sqrt{\mu^\alpha}} c^\alpha_i ] |\varphi_i> .
\end{eqnarray*}
We call such single particle orbits $\{|p>\}$ and energies
$\{\epsilon^p\}$ AMD-HF orbits and energies, respectively 
since they are extracted from AMD wave function.

Here we should note that these orbits and energies of AMD-HF are not 
totally equivalent to those of Hartree-Fock. The reason is as below. 
In AMD+HF, the Hartree-Fock equation is solved only within the
functional space of single particle wave functions which is spanned by 
$\{\varphi_i\}$. Thus in AMD+HF the Hartree-Fock self-consistency is
satisfied only within this restricted functional space. We can regard
the AMD+HF method as being a kind of restricted Hartree-Fock method. 
However this restriction has a large merit because the  functional 
space spanned by the single particle wave functions $\{\varphi_i\}$ is
obtained by the energy variation including parity projection.

\subsubsection{Identification of AMD-HF orbits}

We need to identify the AMD-HF orbits $|p>$ defined in the previous
section. 
As almost nuclei seem to have axially symmetric deformation
approximately, we expect that z component of the total angular
momentum, $\hat{j}_z$, is approximately a good-quantum number, but that the
magnitude of it, $\hat{\bf j}^2$, is not.
In addition, when the states that have the same absolute value of $j_z$ 
are degenerate, the expectation value of $\hat{j}_z$ doesn't give us 
useful information. Therefore we use the square of $\hat{j}_z$
,$\hat{j}_z^2$, so as to avoid this 
difficulty. For example, let us consider the state $|p>$ which
includes $|j_z=m>$ and 
$|j_z=-m>$ as below:
\begin{eqnarray*}
|p> & = & \frac{1}{\sqrt{1+|c|^2}} \; [ \; |m> + c |-m>] .
\end{eqnarray*}
The expectation value of $\hat{j}_z$ is
\begin{eqnarray*}
<p| \hat{j}_z |p> & = & m \cdot \frac{1-|c|^2}{1+|c|^2} .
\end{eqnarray*}
On the other hand, the expectation value of $\hat{j}_z^2$ is
\begin{eqnarray*}
<p| \hat{j}_z^2 |p> & = & m^2 ,
\end{eqnarray*}
from which we can identify {\em m} certainly.

By the way, this identification is done on the body-fixed frame as noticed
in the section about calculating the deformation. Here we calculate 
the inertial axes explicitly by diagonalizing the tensor $Q_{ij}$, which was 
introduced in section II.A.3. We choose the z-axis of the body-fixed frame 
to the direction of the $Q^{space}$'s eigenvector which has the largest 
eigenvalue. Here $Q^{space}$ represents {\em Q} calculated in the
space-fixed frame.
Using the eigenvector ${\bf c}^q$, we can relate $\hat{\bf j}^{space}$ 
and $\hat{\bf j}^{body}$ as follows:
\begin{eqnarray*}
\hat{j}^{body}_q & = & {\bf c}^q \cdot \hat{\bf j}^{space} ,
\end{eqnarray*}
where $\hat{\bf j}^{space}$ and $\hat{\bf j}^{body}$ are the total angular 
momentum operator in space-fixed frame and that in the body-fixed 
frame,
respectively. Thus the square $j_z$ in the body-fixed frame is
represented using the quantities in the space-fixed frame as below:
\begin{eqnarray*}
<p|\hat{j}^2_z|p>_{body-fixed} & = &
\sum_{ij} c^z_i c^z_j <p|\hat{j}_i 
\hat{j}_j |p>_{space-fixed} .
\end{eqnarray*}
In addition, we use also the square of the orbital angular momentum 
$\hat{\bf l}^2$, which is not a good-quantum number. This enables us 
to relate 
the obtained levels to the levels which would be obtained if the shape were 
spherical.
 
\section{AMD-HF orbits in Be isotopes}

In this section we show the results about AMD-HF orbits of the ground states 
that are obtained by AMD. Here AMD  single 
particle wave functions are composed of the single Gaussians.

In TABLE I, the AMD-HF levels are given. $\Omega^2$ and $L^2$  indicate 
the expectation values of the operators $\hat{j}_z^2$ and $\hat{\bf l}^2$ 
respectively, which are obtained by the method mentioned in section II.B.2.
$\beta$ and B.E. are the deformation parameter and the binding energy 
respectively.
We explain the results, about $^9$Be, $^{10}$Be and $^{11}$Be, that look
 interesting. Then we compare these with the simple model 
--- two-center shell model --- to help our understanding.

\vspace{0.5cm}

{\bf a.$^9$Be}

In the minus-parity AMD state of $^9$Be, $\Omega^2$ of the highest level of 
neutron is 1.260, which is not equal to any of
0.25(=$(\frac{1}{2})^2$), 2.25(=$(\frac{3}{2})^2$), etc. 
This means the deformation is not axial symmetric. 
But $L^2$=2.995 implies that 
this level contains rather large amount of the 0p$_{\frac{3}{2}}$
component.

In the plus-parity state, the last neutron is in the level of $\Omega=
\frac{1}{2}$. 
In addition $L^2$ of this level is 4.884. As this value is rather
large, 
we can suppose that this level consists largely of d-orbit. 
In the two center shell model given in FIG.1, the plus-parity level 
named $\gamma$ level in FIG.1 comes down in energy 
when the distance between two alpha clusters is large. This is 
consistent with our result because the deformation parameter $\beta$
obtained in AMD  for the plus-parity state is 0.801, 
which is very large.

\vspace{0.3cm}

\begin{scriptsize}

\twocolumn

\begin{center}
\begin{tabular}{rcccc} 
\multicolumn{5}{l}{{\bf $^{6}$Be(+)} \, B.E.=-21.26 MeV \,
$\beta$=0.437} \\
\hline \hline
s.p. energy & $\Omega^2$ & ${\bf l}^2$ & occ.& plus \\ 
\hline
\multicolumn{5}{l}{proton} \\
$-$2.48  & 0.250 & 2.077 & 2 &  7.4 \\ 
$-$24.47 & 0.250 & 0.118 & 2 & 96.0 \\
\multicolumn{5}{l}{neutron} \\
$-$29.51 & 0.250 & 0.148 & 2 & 93.1 \\
\hline \hline
\\
\\
\\
\multicolumn{5}{l}{{\bf $^{7}$Be($-$)} \, B.E.=$-$30.48 MeV \,
$\beta$=0.566} \\
\hline \hline
s.p. energy & $\Omega^2$ & ${\bf l}^2$ & occ. & plus \\ 
\hline
\multicolumn{5}{l}{proton} \\
$-$11.04 & 0.250 & 2.083 & 2 & 5.1 \\
$-$28.88 & 0.250 & 0.219 & 2 & 96.5 \\
\multicolumn{5}{l}{neutron} \\
$-$11.15 & 0.250 & 2.059 & 1 & 4.3 \\
$-$28.87 & 0.250 & 0.224 & 1 & 96.5 \\
$-$32.04 & 0.250 & 0.440 & 1 & 82.2 \\
\hline \hline
\\
\\
\\
\multicolumn{5}{l}{{\bf $^{8}$Be(+)} \, B.E.=$-$48.00 MeV \, $\beta$=0.656}\\
\hline \hline
s.p. energy & $\Omega^2$ & ${\bf l}^2$ & occ. & plus \\ 
\hline 
\multicolumn{5}{l}{proton} \\
$-$18.99 & 0.250 & 2.189 & 1 & 2.3 \\
$-$19.58 & 0.250 & 2.127 & 1 & 0.2 \\
$-$34.69 & 0.250 & 0.345 & 2 & 100.0 \\
\multicolumn{5}{l}{neutron} \\
$-$18.87 & 0.250 & 2.203 & 1 & 2.7 \\
$-$19.55 & 0.250 & 2.131 & 1 & 0.3 \\
$-$34.69 & 0.250 & 0.345 & 2 & 99.7 \\
\hline \hline
\\
\\
\\
\multicolumn{5}{l}{{\bf $^{9}$Be($-$)} \, B.E.=$-$46.96 MeV \, $\beta$=0.521}\\
\hline \hline
s.p. energy & $\Omega^2$ & ${\bf l}^2$ & occ. & plus \\ 
\hline
\multicolumn{5}{l}{proton} \\
$-$21.82 & 0.249 & 2.098 & 2 & 0.7 \\
$-$39.03 & 0.250 & 0.231 & 2 & 99.7 \\
\multicolumn{5}{l}{neutron} \\
+1.56 & 1.260 & 2.996 & 1 & 21.8 \\
$-$18.18 & 0.269 & 2.147 & 1 & 2.8 \\
$-$21.82 & 0.249 & 2.098 & 1 & 0.7 \\
$-$33.70 & 0.256 & 0.297 & 1 & 99.1 \\
$-$39.03 & 0.249 & 0.230 & 1 & 99.7 \\ 
\hline \hline
\\
\\
\\
\multicolumn{5}{l}{{\bf $^{9}$Be(+)} \, B.E.=$-$42.97 MeV \, 
$\beta$=0.801}\\  
\hline \hline
s.p. energy & $\Omega^2$ & ${\bf l}^2$ & occ. & plus \\ 
\hline
\multicolumn{5}{l}{proton} \\
$-$22.48 & 0.250 & 2.287 & 2 & 3.1 \\
$-$35.04 & 0.250 & 0.494 & 2 & 99.6 \\
\multicolumn{5}{l}{neutron} \\
+6.79 & 0.250 & 4.885 & 1 & 79.8 \\
$-$18.07 & 0.250 & 2.280 & 1 & 4.5 \\
$-$22.37 & 0.250 & 2.320 & 1 & 4.1 \\
$-$31.45 & 0.250 & 0.483 & 1 & 98.6 \\
$-$35.06 & 0.250 & 0.475 & 1 & 99.4 \\
\hline \hline 
\end{tabular}
\end{center}

\begin{center}
\begin{tabular}{rcccc}
\multicolumn{5}{l}{{\bf $^{10}$Be(+)} \, B.E.=$-$53.11 MeV \,
$\beta$=0.381}\\
\hline \hline
s.p. energy & $\Omega^2$ & ${\bf l}^2$ & occ. & plus \\ 
\hline
\multicolumn{5}{l}{proton} \\
$-$23.00 & 0.254 & 2.049 & 2 & 0.5 \\
$-$43.30 & 0.254 & 0.108 & 2 & 99.6 \\
\multicolumn{5}{l}{neutron} \\
$-$7.47 & 1.241 & 2.487 & 2 & 12.7 \\
$-$19.55 & 0.278 & 2.162 & 2 & 3.2 \\
$-$37.72 & 0.257 & 0.100 & 2 & 99.7 \\
\hline \hline
\\
\\
\\
\multicolumn{5}{l}{{\bf $^{10}$Be($-$)} \, B.E.=$-$44.95 MeV \,
$\beta$=0.637}\\
\hline \hline
s.p. energy & $\Omega^2$ & ${\bf l}^2$ & occ. & plus \\ 
\hline
\multicolumn{5}{l}{proton} \\
$-$24.65 & 0.250 & 2.106 & 2 & 0.0 \\
$-$39.75 & 0.250 & 0.304 & 2 & 100.0 \\
\multicolumn{5}{l}{neutron} \\
$-$3.41 & 0.667 & 3.387 & 2 & 59.4 \\
$-$20.39 & 0.264 & 2.328 & 2 & 1.4 \\
$-$35.23 & 0.254 & 0.270 & 2 & 99.7 \\
\hline \hline
\\
\\
\\
\multicolumn{5}{l}{{\bf $^{11}$Be(+)} \, B.E.=$-$47.25 MeV \,
$\beta$=0.505}\\
\hline \hline
s.p. energy & $\Omega^2$ & ${\bf l}^2$ & occ. & plus \\  
\hline
\multicolumn{5}{l}{proton} \\
$-$25.97 & 0.250 & 2.089 & 2 & 1.9 \\
$-$44.97 & 0.250 & 0.140 & 2 & 99.4 \\
\multicolumn{5}{l}{neutron} \\
+4.25 & 0.272 & 4.463 & 1 & 97.1 \\
$-$6.74 & 1.273 & 2.153 & 1 & 3.7 \\
$-$8.84 & 1.253 & 2.543 & 1 & 12.3 \\
$-$17.75 & 0.250 & 2.327 & 1 & 2.5 \\
$-$22.47 & 0.258 & 2.264 & 1 & 6.2 \\
$-$35.43 & 0.252 & 0.147 & 1 & 99.8 \\
$-$39.55 & 0.255 & 0.134 & 1 & 99.8 \\
\hline \hline
\\
\\
\\
\multicolumn{5}{l}{{\bf $^{11}$Be($-$)} \, B.E.=$-$54.70 MeV \,
$\beta$=0.271}\\
\hline \hline
s.p. energy & $\Omega^2$ & ${\bf l}^2$ & occ. & plus \\ 
\hline
\multicolumn{5}{l}{proton} \\
$-$24.97 & 0.250 & 2.034 & 2 & 1.2 \\
$-$48.62 & 0.250 & 0.042 & 2 & 99.7 \\
\multicolumn{5}{l}{neutron} \\
$-$2.52 & 1.314 & 2.350 & 1 & 8.2 \\
$-$7.15 & 1.241 & 2.315 & 1 & 7.5 \\
$-$10.75 & 1.388 & 2.311 & 1 & 7.4 \\
$-$18.04 & 0.251 & 2.144 & 1 & 4.7 \\
$-$21.55 & 0.259 & 2.115 & 1 & 2.9 \\
$-$36.61 & 0.250 & 0.051 & 1 & 99.7 \\
$-$42.60 & 0.253 & 0.035 & 1 & 100.0 \\
\hline \hline
\\
\end{tabular}
\end{center}

TABLE I. Single particle levels of the ground state of each
nucleus. ''B.E.'' and ''$\beta$'' are the binding energy and the
deformation parameter, respectively. ''occ.'' indicates the occupation 
number of the
level. ''plus'' indicates the percentage of plus-parity
component. We regard two levels as the same
level whose energy difference is less than 100 keV. In such case, the
difference in $\Omega^2$ and ${\bf l}^2$ is about 0.01.

\end{scriptsize}

\onecolumn

{\bf b.$^{10}$Be}

The calculated plus-parity state of $^{10}$Be has a good correspondence 
with the two-center shell 
model, if we think our highest level of neutron corresponds to the
level named the $\alpha$ level in FIG.1 which comes from 
$j_z$=$\frac{3}{2}$ of 0p$_\frac{3}{2}$ in spherical case.

But on the other hand, the minus-parity state does not have a
similarity with the two-center shell model. 
The most surprising point is that the only three spectra 
of neutron appeared in our calculation. In the two-center shell model 
every single particle level is the eigenstate 
of parity. So in the model four spectra should appear. 
The explanation of this result by AMD+HF is as below:
`` The single particle orbit (AMD-HF orbit) extracted from AMD wave
function is 
the parity-mixing state. And the last two neutrons are degenerate 
in that state.''
Explaining this phenomenon in terms of the two-center shell model is
as follows. 
At some distance between two $\alpha$ particles, the levels $\Omega=
\frac{3}{2}$ and $\Omega=\frac{1}{2}$, coming from 0p$_\frac{3}{2}$ and
0d$_\frac{5}{2}$ respectively, approach each other. As a result, these levels 
are mixed, and then we see the only one level in which two levels are
 mixed. 
Here we investigate the percentage of parity-mixing in the last AMD-HF
orbit of neutron. This AMD-HF level, which is represented as $|f>$, can
be written as follows:
\begin{eqnarray*}
|f> & = & \epsilon_1|+> + \epsilon_2|-> ,\\
& & \epsilon_1^2+\epsilon_2^2=1 ,
\end{eqnarray*}
where $|+>$ and $|->$ indicate plus- and minus-parity components,
 respectively. 
Using the operator ${\cal P}$ which reverses the parity, we calculate the 
following quantity A:
\begin{eqnarray*}
A & \equiv & <f|{\cal P}|f> = \epsilon_1^2 - \epsilon_2^2 .
\end{eqnarray*}
Then we obtain the ratio as follows:
\begin{eqnarray*}
\epsilon_1^2 = \frac{1+A}{2} \; , \; \epsilon_2^2 = \frac{1-A}{2} . 
\end{eqnarray*}
Calculated results are $\epsilon_1^2$=0.60 
and $\epsilon_2^2$=0.40. As we expected this level is a state with
large parity-mixing.

\vspace{0.3cm}

{\bf c.$^{11}$Be}

In the calculated plus-parity state of $^{11}$Be, the last neutron
occupies what we call ''halo level''. 
$\Omega^2$=0.25 means $\Omega$=$\frac{1}{2}$ and $L^2$=4.463 of this
level means that this level
contains a large component of  
0d$_\frac{5}{2}$. We think that  this result is similar to that of the 
two-center shell model with medium distance between two $\alpha$
particles. The obtained value of the deformation parameter 
$\beta$=0.504, which is medium, is consistent with the medium distance 
in the two-center shell model.

However in the calculated minus-parity state of $^{11}$Be, 
we don't find so good
similarity between AMD+HF and the two-center shell model. 
As $\beta$=0.271 is rather small, we try to compare the 
obtained result to that of the two-center shell model with small
inter-$\alpha$ distance. 
In that case, the last neutron should be put into the level named the
$\beta$ level in FIG.1 which has $\Omega$=$\frac{1}{2}$ and 
comes from 0p$_\frac{1}{2}$ in spherical case. But according to AMD+HF 
calculation, $\Omega^2$ of the last neutron level is 1.314, which means
that $\Omega$ is not $\frac{1}{2}$.

By the way, there is the famous problem in $^{11}$Be, which is that the parity 
of the ground state is not minus but plus. In this calculation, we have 
not succeeded to reproduce this anomalous parity just like the
previous AMD calculation of Ref.\CITE{AMD}. We think that this
failure comes from the insufficient description of the single particle 
orbits in the present calculation where we have assigned only one
Gaussian for one nucleon.
 
\section{AMD-HF orbits with constrained deformation}

In this section, we show the AMD-HF orbits at various $\beta$, which look like 
Nilsson diagram. The levels at each $\beta$ are calculated by the method as 
mentioned in the section II.A.2. As in the previous section, we show only the 
results about $^9$Be($-$), $^{10}$Be($-$) and $^{11}$Be($-$), 
which look 
interesting. We show the diagram of the neutron's AMD-HF levels of
$^9$Be($-$), $^{10}$Be($-$) and $^{11}$Be($-$) in FIG.2(a), FIG.3(a) and
FIG.4(a), respectively. As 
the behavior of the highest level in each diagram looks especially 
interesting, we show the variation
of $L^2$, $\Omega^2$ and the percentage of the plus-parity component of 
the highest level in FIG.2(b), FIG.3(b) and FIG.4(b). 
In addition as the second 
and the third levels in $^{11}$Be($-$) look interesting too, we show the 
same quantities about these levels in FIG.4(c) and FIG.4(d). 
Notice that the scale on the right hand side is used for the
percentage of the plus-parity component.

\vspace{0.5cm}

{\bf a.$^9$Be($-$)}

\begin{figure}[h]
\epsfxsize=0.6 \textwidth
\centerline{\epsffile{ 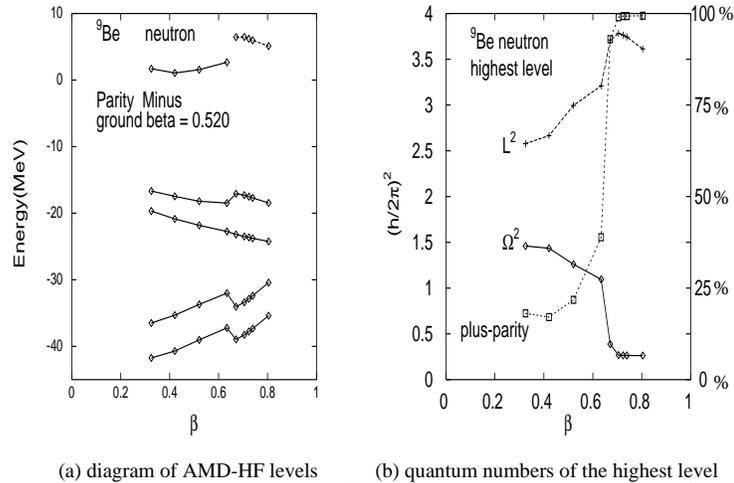}} 
\caption{(a) Diagram of the AMD-HF levels of $^9$Be($-$). (b)
Variation of $L^2$, $\Omega^2$ and
the percentage of the
plus-parity component of the highest level.}
\end{figure}

In FIG.2(a) we can see the inversion or the crossing of the single particle 
levels like that appearing 
in the two-center shell model. When we trace the behavior of the last
neutron's level, in FIG.2(b) we see that up to $\beta$=0.67 the last
level seems to be $|\Omega|=\frac{3}{2}$ coming from 
0p$_\frac{3}{2}$ in spherical case. But beyond $\beta$=0.67, 
its behavior changes apparently. Suddenly $L^2$ becomes rather large,
$\Omega^2$ becomes about 0.25 and the percentage of the plus-parity 
component becomes about 100\%. Therefore we can recognize it as the
lowered halo level that is 
$|\Omega|=\frac{1}{2}$ coming from 0d$_\frac{5}{2}$. 
We notice that the behaviors of the other levels than the highest
level are 
also changed beyond $\beta$=0.67 in FIG.2(a), but that 
$L^2$,$\Omega^2$ and the 
percentage of the plus-parity component are not changed. 
As the parity of the last neutron is almost plus beyond 
$\beta$=0.67, the total intrinsic wave function is an almost
plus-parity state. Therefore the projection to the minus-parity state 
is made by picking up the very small component of minus parity.

\vspace{0.3cm}

{\bf b.$^{10}$Be($-$)}

In FIG.3(a), we can see that the last neutron's level keeps to be 
parity-mixed even if $\beta$ is changed. When we look at the behavior
of the quantum numbers $\Omega$ and {\em L} in FIG.3(b), we find that 
as $\beta$ becomes larger $|\Omega|$ approaches to $\frac{1}{2}$ and
{\em L} increases too. 
This result seems to indicate that this level has two components, 
one being $|\Omega|=\frac{1}{2}$ coming from 0d$_\frac{5}{2}$ and 
the other being $|\Omega|=\frac{3}{2}$ coming 
from 0p$_\frac{3}{2}$, and that as $\beta$ becomes larger the component of 
$|\Omega|=\frac{1}{2}$ increases. 
In addition, according to the parity-mixing ratio, this insight seems to be
 correct, 
since as $\beta$ becomes larger the ratio of plus-parity component increases.

\begin{figure}[h]
\epsfxsize=0.6 \textwidth
\centerline{\epsffile{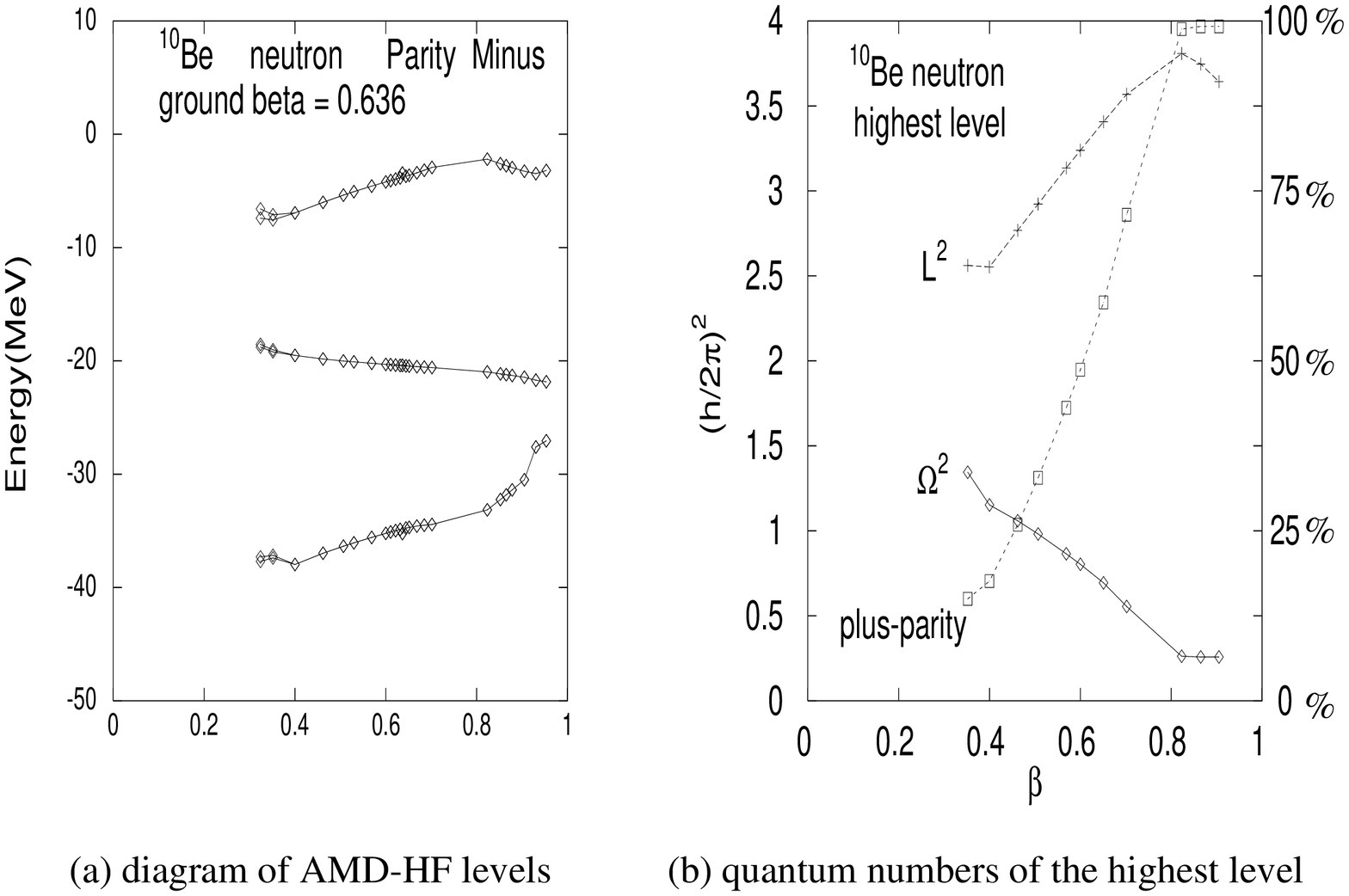}}
\caption{(a) Diagram of the AMD-HF levels of $^{10}$Be($-$). (b)
Variation of $L^2$, $\Omega^2$ and
the percentage of the 
plus-parity component of the highest level.}
\end{figure}

\vspace{0.3cm}

{\bf c.$^{11}$Be($-$)}

We can see the tendency that the halo level, which is 
$|\Omega|=\frac{1}{2}$ coming 
from 0d$_\frac{5}{2}$, is lowered at large $\beta$. Especially at 
$\beta$=0.7 $\sim$ 0.8 the tendency is prominent. In that region, 
we can see $L^2 \sim 2$, $\Omega^2 \sim 2$ and the parity is 100\%
minus for the highest level in FIG.4(b). And in FIG.4(c) and 4(d), $L^2$ is 
rather large, $\Omega^2 \sim 0.25$ and the 
parity is about 90\% plus for the second and third levels. 
Therefore we can identify in this region the highest level as 
$|\Omega|=\frac{3}{2}$ coming 
from 0p$_\frac{3}{2}$ and the second and third levels as 
$|\Omega|=\frac{1}{2}$ coming from 0d$_\frac{5}{2}$ which is the 
lowered halo level. 
Note that these two halo levels are not degenerate in this 
odd-neutron-number nucleus $^{11}$Be. 
But beyond $\beta$=0.8 the properties of the AMD-HF levels change
 again. 
Especially in FIG.4(b) (the last neutron), the plus-parity component 
increases and $L^2$ becomes very large. We can explain this phenomenon with 
the two-center shell model as follows. The large deformation parameter $\beta$ 
corresponds to the large inter-$\alpha$ distance in the two-center 
shell model. 
As shown in FIG.1, in such case the $\alpha$ level and $\gamma$ level approach 
very closely and are degenerate. In our calculation the same phenomenon
as the two-center shell
model seems to happen and then the 
component of the halo orbit 
mixes into the highest level. 

\begin{figure}[h]
\epsfxsize=1.0 \textwidth
\centerline{\epsffile{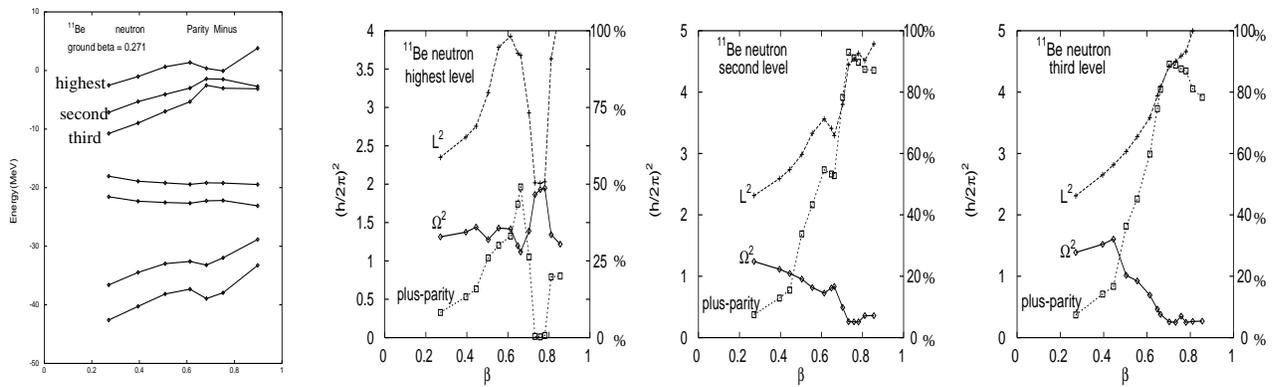}}
\caption{(a) Diagram of the AMD-HF levels of $^{11}$Be($-$). (b)
 Variation of $L^2$, $\Omega^2$ and
the percentage of the
plus-parity component of the highest level. (c) and (d) The same
quantities as (b) for the
second and the third level, respectively.}
\end{figure}

\vspace{1cm}

\section{Excited positive parity band in $^{10}$Be}

The original AMD does not have the concept of the
single particle orbit in
the mean field. However we have confirmed in sections III and IV 
that AMD contains in it
the mean field (and its single particle orbits). 
In sections III and IV, we used the single particle orbits only for 
''analyzing the AMD wave function''. In this section, we
truly ''use'' the single particle orbits by performing ''particle-hole 
excitation'' for the ground state. 
Our idea is that if the AMD-HF orbits are really meaningful, the
description of excited states by the particle-hole excitation should
be meaningful.

Here we apply such p-h excitation for constructing the first excited $0^+$ 
state of $^{10}$Be at 6.179 MeV which is just above the lowest 1$^-$
state at 5.960 MeV\CITE{10Be}. One of possible configurations for 
this 0$^+$ is 
that two neutrons occupy the halo orbit, which is
$\Omega$=$\frac{1}{2}$ level 
coming from 0d$_\frac{5}{2}$ in the two-center shell model. 
So we want to 
confirm that the second $0^+$ constructed by such p-h excitation appears 
in the vicinity of the lowest 1$^-$ as shown by the experiment.

According to the two-center shell model, 
the lowest minus-parity state, which has the total angular momentum 1, 
seems 
to have the configuration that each of the
levels, $\Omega^\pi$=$\frac{3}{2}^-$ from 0p$_\frac{3}{2}$($\alpha$ level)
and $\Omega^\pi$=$\frac{1}{2}^+$ from 0d$_\frac{5}{2}$($\gamma$ level) 
which is ''halo level'', is occupied by one neutron. Therefore by 
raising up the 
neutron occupying $\Omega^\pi$=$\frac{3}{2}^-$ to 
$\Omega^\pi$=$\frac{1}{2}^+$, the 
configuration becomes $[\Omega^\pi=\frac{1}{2}^+]^2$ which gives us 
the first excited plus-parity state having the total angular momentum 0.

But according to the AMD-HF calculation of the lowest minus-parity state, 
we don't have good parity $\Omega^\pi$=$\frac{3}{2}^-$ and 
$\Omega^\pi$=$\frac{1}{2}^+$ levels but one parity-mixing level. 
In addition, the levels except 
this one are eigenstates of parity. Therefore in this case we can easily make
the first excited plus-parity state from the lowest minus-parity state 
only by 
projecting out from the intrinsic state of the lowest minus-parity 
state the plus-parity state: i.e. when the lowest minus-parity state 
$|-_g>$ is represented with its intrinsic state $|\Phi_{int}>$ as 
\begin{eqnarray*}
|-_g> & = & \frac{1}{\sqrt{2}} [|\Phi_{int}> - {\cal P}|\Phi_{int}>],
\end{eqnarray*}
the first excited plus-parity state $|+'_1>$ is formed as
\begin{eqnarray*}
|+'_1> & = & \frac{1}{\sqrt{2}} [|\Phi_{int}> + {\cal P}|\Phi_{int}>] .
\end{eqnarray*}
But this $|+'_1>$ may not be orthogonal to the lowest plus-parity 
state $|+_g>$ whose value 
of $\beta$ is 0.381. In order to avoid the mixing of this lowest plus-parity  
state, we orthogonalize $|+'_1>$ to $|+_g>$ as below:
\begin{eqnarray*}
|+_1> & \equiv & |+'_1> - \frac{<+'_1|+_g>}{<+_g|+_g>} \cdot |+_g> .
\end{eqnarray*}

Though $|+_1>$ and $|\pm_g>$ are not the eigenstates of the total 
angular momentum, we can calculate 
the approximate energies of the projected states as below. For a given
intrinsic state $|\Phi>$, the approximate expression for the energy
$E_I$ of the projected state with angular momentum {\em I} is
\begin{eqnarray*}
E_I & = & <\Phi|\hat{H}|\Phi> - \frac{\hbar^2}{2 {\cal J}_x}
<\Phi|\hat{\bf J}^2
|\Phi> + \frac{\hbar^2}{2 {\cal J}_x}I(I+1) ,\\
& &  {\cal J}_x \equiv <\Phi|\sum^A_{i=1} (\hat{y}^2_i
+\hat{z}^2_i)|\Phi> .
\end{eqnarray*}

\begin{figure}[h]
\epsfxsize=0.66 \textwidth
\centerline{\epsffile{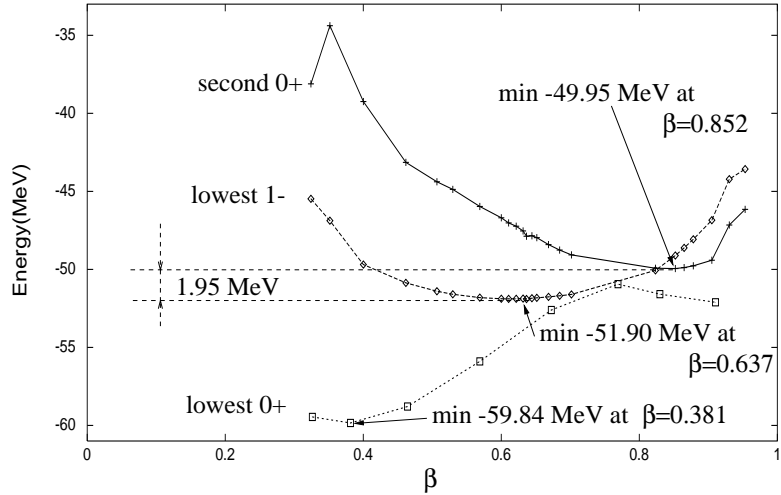}}
\caption{Projected energies of $^{10}$Be states as functions of $\beta$}
\end{figure}

In FIG.5, we show the projected energies of $^{10}$Be states as functions
of $\beta$. 
According to FIG.5, the minimum energy and $\beta$ of each state is as below.
\begin{eqnarray*}
{\rm lowest} \; 0^+ & : & -59.84 \;{\rm MeV \; at} \; \beta=0.381 \\
{\rm lowest} \; 1^- & : & -51.90 \;{\rm MeV \; at} \; \beta=0.637 \\
{\rm second}  \; 0^+ & : & -49.95 \;{\rm MeV \; at} \; \beta=0.852 
\end{eqnarray*}
Therefore we can see that the first excited 0$^+$ state appears in the 
 vicinity of the lowest 1$^-$ state with the small energy difference of 1.95 
MeV.

As mentioned previously, the AMD wave functions of 
Be isotopes have the core part with an approximate
dumbbell structure of two alpha clusters. Therefore 
the inter-alpha distance is also useful as a measure of the
deformation of Be isotopes. The above three values of $\beta$ for
$0^+_1$, $1^-$, and $0^+_2$ in $^{10}$Be correspond to the
inter-alpha distance, 1.99 fm, 2.66 fm, and 3.55 fm, respectively.

\section{Summarizing discussion}

In this paper we have proposed a new theoretical method for the study
of nuclear structure, which we have called the ''AMD + Hartree-Fock''
method (AMD+HF). The purpose of introducing this new method is to
develop the AMD approach by incorporating into it the concept of the
single particle motion in the mean field. The new method consists of
the following two steps. In the first step we construct the AMD wave
function by using usual AMD method. Then in the next step we extract
the physical single particle motion in the mean field which is
contained in the AMD wave function. This is done by diagonalizing the
Hartree-Fock Hamiltonian in the functional space spanned by the single 
particle wave functions which constitute the AMD wave function. In this 
paper, for the sake of simplicity, we represented the AMD single
particle wave function by the single Gaussian wave packet. But for the
better description of the single particle motion we had better to 
represent the AMD single
particle wave function by the superposition of
Gaussian wave packets. Once we obtain the physical single particle wave 
functions in the mean field, we can utilize them for many
purposes. First we can use them for understanding the physical
situation of the ground state (or the ground rotational band states)
of the system which the AMD wave function describes. Second we can use them 
for constructing the wave functions for the excited states of the
system by constituting the particle-hole configurations. The latter way 
of utilizing the physical
single particle wave functions in the mean field is
beyond the scope of the ordinary AMD approach.

The AMD+HF method is applicable for any nuclei, but it is expected to
be especially useful for the study of unstable nuclei. The reason is as 
follows. The original AMD itself has proved to be very powerful for the 
study of unstable nuclei, which has been largely due to its basic
character that it needs no model assumptions such as the axial
symmetric deformation and the existence of the clusters. However, this
merit of the original AMD means at the same time that we need to
perform adequate analyses of the obtained AMD wave function in order
to clarify the physics which the AMD wave function contains. As an
example let us consider Be isotopes which we have treated in this
paper. According to the AMD study of Be isotopes, the wave function of Be
isotopes have the core part with an approximate dumbbell structure of
 two alpha clusters even for very neutron-rich isotopes near
neutron dripline. Then there naturally arise questions about what kind
of dynamics governs so many neutrons distributed around the core part
and about what kinds of interaction are existent between the core part 
and the neutrons. Our new approach, AMD+HF, can give us important
information in answering to these questions.

It should be noted that the AMD+HF method inherits most of the
advantageous features of the original AMD. We need no model
assumptions, we can perform variational calculation after projecting
parity, we can use realistic effective nuclear force with finite
range, we can perform angular momentum projection rather easily, we can
superpose Slater determinants with no difficulty, and so on. Here we
make a comment about the superposition of Slater determinants. In
performing the ordinary AMD calculation, the superposition of Slater
determinants causes no problem as was reported in previous
papers\CITE{AMD}. What
we need to comment here is about the extraction of the single particle 
wave functions in the mean field contained in the superposed Slater
determinants. In this paper, we have explained how to extract the mean
field only in the case where the AMD wave function is expressed by a
single Slater determinant. In the general case of the AMD wave function 
given by superposed Slater determinants, we first calculate the single
particle density matrix from AMD wave function. Then we can calculate
the Hartree-Fock type single particle Hamiltonian by the use of the
density matrix. We will discuss this generalization elsewhere.

As was pointed out in previous papers, the AMD method which adopts a
single Gaussian wave packet for the single particle wave function is
not  suitable for the description of the long tails of the neutron
halo orbits. For halo phenomena we have to adopt a superposition of
Gaussian wave packets for the single particle wave function. However to 
represent all the single particle wave functions by the superposition
of many Gaussian wave packets means a very heavy computational
work. One of the aims of the AMD+HF method is to make more adequate
description of the long tails of neutron halo orbits than the ordinary 
AMD method. By using the AMD+HF method we can identify the least bound
neutron orbits which correspond to neutron halo orbits. What is
necessary for us is to improve the single particle wave functions only 
for these least bound neutron orbits. Thus we expect that the AMD+HF
method will enable us to treat neutron halo orbits in more efficient
way than the usual AMD method. We will discuss this problem elsewhere.

Generally speaking, the deformation of excited states is different
from that of the ground state. Therefore in order to construct the
wave functions with particle-hole excited configurations, we have to
prepare the single particle wave functions for various magnitudes of
quadrupole deformation. In this paper, we first calculated the
minimum-energy AMD wave functions for various magnitudes of
deformation by the use of the frictional cooling method under the
constraint of the deformation and then extracted the single particle
orbits from the obtained AMD wave functions. The investigation of the
properties of the obtained single particle wave functions including
the deformation-dependence of the single particle energies have shown
that these AMD-HF orbits as functions of quadrupole deformation are
rather similar to the single particle orbits of the two-center shell
model quoted in section I. In spite of the overall similarity between
AMD-HF orbits and two-center shell model orbits, there exist some
interesting differences. Among them the appearance of the parity-mixed 
AMD-HF orbits is remarkable since it is out of the scope of the
two-center shell model. A good example of the parity-mixed AMD-HF
orbit is the least bound orbit extracted from the minus-parity AMD
wave function of $^{10}$Be. This orbit comes down in energy for large
deformation and can be regarded as corresponding to the ''halo orbit''.

In this paper we have studied Be isotopes by our new AMD+HF
method. However the application of the AMD+HF method to the study of
excited states by constructing the wave functions with particle-hole
excited configurations has been made only for one problem, namely the
study of the second 0$^+$ state of $^{10}$Be at 6.18 MeV. According to
the calculation reported in this paper, the 0$^+$ state with the
configuration of the two valence neutrons occupying the valence halo
orbit which is parity mixed have the excitation energy near that of
the lowest 1$^-$ state. 
This result is consistent with the experimental data which show that 
the second 0$^+$ state at 6.18 MeV exists in the vicinity of the
lowest 1$^-$ state at 5.96 MeV.
We here would like to comment that this calculated result can be
improved by using the
density-dependent force. The density-dependent force works
attractively at low density, and repulsively at high
density. Therefore in the case where the density-dependent force is 
added to the 
force used in this paper, the second 0$^+$ state comes down in energy 
more than the
lowest 1$^-$ state does, because the second 0$^+$ state is deformed
more largely than the lowest 1$^-$ state according to the calculated result. 
We expect that the second 0$^+$ state appears more closely to the
lowest 1$^-$ state.

Our study of Be isotopes in this paper was made
for the region of $^6$Be $\sim$ $^{11}$Be. But of
course we can study $^{12}$Be and  $^{14}$Be
in the same way. Especially it is very interesting to study 
the role of the ''halo orbit'' not only in the ground state but also in
the excited states in these $^{12}$Be and $^{14}$Be. 

In summary, the AMD-HF theory can be expected to be not only useful
for the analysis of the AMD wave function but also powerful for the
systematic study of excited states in a different way from the
ordinary AMD approach by relying on the concept of the particle-hole
excitation.

\section{Acknowledgements}

The authors thank Prof. von Oertzen for
many valuable discussions about Be isotopes and
molecular orbital description for them. One 
of the authors (A. D.) thanks Mr. H. Takemoto,
Mr. T. Fujita, and
Mr. H. Shin for discussions.

\end{document}